 \documentclass{emulateapj}

\slugcomment{To appear in ApJ main journal.}

\shorttitle{CANGAROO-III Observations of RX J0852.0$-$4622}
\shortauthors{Enomoto et al.}

\begin{document}

\title{CANGAROO-III Observations of the supernova remnant
RX J0852.0$-$4622}

\author{
R.~Enomoto\altaffilmark{1}
S.~Watanabe\altaffilmark{2}
T.~Tanimori\altaffilmark{2}
A.~Asahara\altaffilmark{2}
G.~V.~Bicknell\altaffilmark{3}
R.~W.~Clay\altaffilmark{4}
P.~G.~Edwards\altaffilmark{5}
S.~Gunji\altaffilmark{6}
S.~Hara\altaffilmark{7}
T.~Hattori\altaffilmark{8}
S.~Hayashi\altaffilmark{9}
Y.~Higashi\altaffilmark{2}
R.~Inoue\altaffilmark{8}
C.~Itoh\altaffilmark{10}
S.~Kabuki\altaffilmark{2}
F.~Kajino\altaffilmark{9}
H.~Katagiri\altaffilmark{11}
A.~Kawachi\altaffilmark{8}
S.~Kawasaki\altaffilmark{1}
T.~Kifune\altaffilmark{1}
R.~Kiuchi\altaffilmark{1}
K.~Konno\altaffilmark{6}
L.~Ksenofontov\altaffilmark{1}
H.~Kubo\altaffilmark{2}
J.~Kushida\altaffilmark{8}
Y.~Matsubara\altaffilmark{12}
T.~Mizukami\altaffilmark{2}
R.~Mizuniwa\altaffilmark{8}
M.~Mori\altaffilmark{1}
H.~Muraishi\altaffilmark{13}
T.~Naito\altaffilmark{14}
T.~Nakamori\altaffilmark{2}
D.~Nishida\altaffilmark{2}
K.~Nishijima\altaffilmark{8}
M.~Ohishi\altaffilmark{1}
Y.~Sakamoto\altaffilmark{8}
V.~Stamatescu\altaffilmark{4}
S.~Suzuki\altaffilmark{15}
T.~Suzuki\altaffilmark{15}
D.~L.~Swaby\altaffilmark{4}
H.~Tanimura\altaffilmark{2}
G.~Thornton\altaffilmark{4}
F.~Tokanai\altaffilmark{6}
K.~Tsuchiya\altaffilmark{2}
Y.~Yamada\altaffilmark{9}
M.~Yamazaki\altaffilmark{9}
S.~Yanagita\altaffilmark{15}
T.~Yoshida\altaffilmark{15}
T.~Yoshikoshi\altaffilmark{1}
M.~Yuasa\altaffilmark{1}
Y.~Yukawa\altaffilmark{1}
}

\altaffiltext{1}{ Institute for Cosmic Ray Research, University of Tokyo, Kashiwa, Chiba 277-8582, Japan} 
\altaffiltext{2}{ Department of Physics, Graduate School of Science, Kyoto University, Sakyo-ku, Kyoto 606-8502, Japan} 
\altaffiltext{3}{ Research School of Astronomy and Astrophysics, Australian National University, ACT 2611, Australia} 
\altaffiltext{4}{ Department of Physics and Mathematical Physics, University of Adelaide, SA 5005, Australia} 
\altaffiltext{5}{ Paul Wild Observatory, CSIRO Australia Telescope National Facility, Locked Bag 194, Narrabri, NSW 2390, Australia} 
\altaffiltext{6}{ Department of Physics, Yamagata University, Yamagata, Yamagata 990-8560, Japan} 
\altaffiltext{7}{ Ibaraki Prefectural University of Health Sciences, Ami, Ibaraki 300-0394, Japan} 
\altaffiltext{8}{ Department of Physics, Tokai University, Hiratsuka, Kanagawa 259-1292, Japan} 
\altaffiltext{9}{ Department of Physics, Konan University, Kobe, Hyogo 658-8501, Japan} 
\altaffiltext{10}{ National Institute of Radiological Sciences, Chiba, Chiba 263-8555, Japan}
\altaffiltext{11}{ Department of Physical Science, Graduate School of Science, Hiroshima University, Higashi-Hiroshima, Hiroshima 739-8526, Japan} 
\altaffiltext{12}{ Solar-Terrestrial Environment Laboratory,  Nagoya University, Nagoya, Aichi 464-8602, Japan} 
\altaffiltext{13}{ School of Allied Health Sciences, Kitasato University, Sagamihara, Kanagawa 228-8555, Japan} 
\altaffiltext{14}{ Faculty of Management Information, Yamanashi Gakuin University, Kofu, Yamanashi 400-8575, Japan} 
\altaffiltext{15}{ Faculty of Science, Ibaraki University, Mito, Ibaraki 310-8512, Japan} 

\begin{abstract}

Sub-TeV gamma-ray emission from the North-West rim of the supernova remnant
RX J0852.0$-$4622 was detected with the CANGAROO-II telescope and
recently confirmed by the H.E.S.S.\ group.
In addition, the H.E.S.S.\ data revealed a very wide 
(up to two degrees in diameter),
shell-like profile of the gamma-ray emission.
We have 
carried out 
CANGAROO-III 
observations in January and February 2005 with three telescopes and
show here the results of three-fold coincidence data.
We confirm the H.E.S.S.\ results 
about the morphology and the energy spectrum,
and find the energy spectrum in the NW-rim is  
consistent with that of the whole remnant.

\end{abstract}


\section{Introduction}

RX J0852.0$-$4622 (G266.2$-$1.2) 
is one of the few supernova remnants (SNRs)
from which strong non-thermal X-ray emission has been detected
\citep{aschenbach,tsunemi,slane}. 
A distance of $\sim$200\,pc and an age of
$\sim$700~years \citep{aschenbach99,tsunemi} would make 
RX J0852.0$-$4622 one of the closest young SNR
\citep[although see][for a dissenting view]{slane}. 
The non-thermal emission suggests particle
acceleration to $\sim$100\,TeV, with the possibility of
TeV gamma-ray emission from inverse Compton scattering
off electrons, and/or $\pi^0$ decay from proton interactions,
with clear implications for the origin of cosmic rays
in our galaxy \citep{volk05}.

Sub-TeV gamma-rays from the northwest rim (NW rim) of RX J0852.0$-$4622 
were detected with the 10\,m CANGAROO-II imaging atmospheric
Cherenkov telescope (IACT) \citep{katagiri}.
Gamma-ray emission was recently confirmed by the
H.E.S.S.\ group \citep{hess0852}, who mapped the spatial distribution
of gamma-ray emission over the whole remnant,
using data from at least two of four telescopes in their array.

RX J0852.0$-$4622 is the second spatially resolved SNR
at TeV gamma-ray energies \citep{enomoto_nature,hess_nature}.
The CANGAROO-II result indicated a soft energy spectrum
$\propto E^{-4.3^{+1.7}_{-4.4}}$, however, in contrast
the H.E.S.S.\ group reported a harder spectrum 
$E^{-2.1\pm0.1\pm0.2}$.
The gamma-ray fluxes also appeared discrepant: at 1\,TeV 
the CANGAROO-II flux level was 15\% of the Crab nebula, whereas the
H.E.S.S.\ result was 100\% of the Crab.
The clear differences between these two reports are;
\begin{itemize}
  \item CANGAROO-II was only sensitive to within 0.4$^\circ$ of the
    X-ray maximum in the NW rim \citep{aschenbach,tsunemi,slane}
    whereas H.E.S.S.\ was sensitive to the whole remnant.
  \item the CANGAROO-II observation used a single telescope whereas
    the H.E.S.S.\ observations used stereo data from two or more telescopes.
  \item the CANGAROO-II flux was derived based on the subtraction of
    a background derived from the surrounding area which included part 
    of the remnant, whereas the H.E.S.S.\ background was derived 
    from an off-source region outside the remnant.
    The CANGAROO-II result, therefore, was dependent on the spatial 
    distribution of gamma-rays inside the SNR.
\end{itemize}
The H.E.S.S.\ result (from stereo observations) showed 
very extended emission, up to one degree in radius
from the center of the remnant.
In this paper we present the results of CANGAROO-III stereoscopic
observations and determine how much of the apparent discrepancies
between these two earlier results arise from these differences.

\section{CANGAROO-III Stereoscopic System}

The use of imaging atmospheric Cherenkov telescopes (IACTs) was
established with the statistically unassailable
detection of the Crab nebula at TeV energies by the
Whipple group \citep{whipple}.
This technique enables TeV gamma-rays to be selected from the
huge background of cosmic rays with the use of the 
``Hillas parameters" of the shower images \citep{hillas}.
Stereoscopic observations, which allow the signal-to-noise ratio
to be significantly improved, were pioneered by the HEGRA group \citep{hegra}.
The H.E.S.S.\ group has recently reported the detection of 
faint gamma-ray sources with an angular
resolution as fine as a few arc-minutes \citep{HESS_science}.

CANGAROO-III is one of two major IACTs located in the southern
hemisphere.
The CANGAROO-III stereoscopic system consists of four imaging atmospheric
Cherenkov telescopes located near Woomera, South Australia (31$^\circ$S,
137$^\circ$E).
Each telescope has a reflector of 10\,m diameter.
Each reflector consists of 114 segmented spherical mirrors (80\,cm
in diameter with a radius of curvature of 16.4\,m) made of FRP
\citep{kawachi} mounted on a parabolic
frame ($f/d$=0.77, i.e., a focal length of 8\,m).
The total light collection area is 57.3\,m$^2$.
The first telescope, T1, which was the CANGAROO-II telescope,
is not presently in use due to its smaller field of view
and higher energy threshold.
The second, third, and fourth telescopes (T2, T3, and T4) were used for the
observations described here.
The camera systems for T2, T3, and T4 are identical and their details
are given in \citet{kabuki}.
The telescopes are located at the 
east (T1), west (T2), south (T3) and north (T4)
corners of a diamond 
with sides of $\sim$100\,m \citep{enomoto_app}.

\section{Observations}

The observations were carried out 
in the period from 2005 January 16 to February 23
using ``wobble mode"
in which the pointing position of each telescope was
shifted in declination between $\pm$0.5 degree from
the center of the remnant (RA, dec = 133.00$^\circ$, $-$46.37$^\circ$
J2000) 
every 20 minutes \citep{wobble}.
Data were recorded for T2, T3 and T4 when
more than four photomultiplier (PMT) signals 
exceeded 7.6 photoelectrons (p.e.) 
in any two telescopes (``Global Trigger") \citep{nishijima}.
The typical trigger rate was 9\,Hz for three-fold coincidence.
Each night was divided into two or three periods, i.e., ON--OFF,
OFF--ON--OFF, or OFF--ON observations. ON-source observations were timed
to contain the meridian passage of the target. 
On average the OFF source regions were located with an offset in RA of 
+30$^\circ$ or $-$30$^\circ$ from
the center of the remnant. The OFF-source observations were also made in
wobble mode.
The total observation time was 1736 and 1408~min, for ON and OFF 
observations, respectively.
Next we required the images in all three telescopes to have clusters
of at least five adjacent pixels exceeding a 5\,p.e.\ threshold
(three-fold coincidence).
The event rate was reduced to $\sim$6\,Hz by this criterion.
Looking at the time dependence of these rates, we can remove data
taken in cloudy
conditions. This procedure is the same as the ``cloud cut''
used in the CANGAROO-II analysis \citep{enomoto_nature}.
We also rejected data taken at elevation angles less than 60$^\circ$.
In total, 1081 min.\ data survived these cuts for ON and
1031 min.\ for OFF, with a
mean elevation angle of 70.3$^\circ$.
The size of this supernova remnant is large (2 degree) even compared to
the field of view (FOV) of our observation (4 degree).
One of the reason why we took "wobble" observation is to enlarge
the effective FOV, the other is to average the responses of 
individual pixels.
We, therefore, took LONG OFF source run of ''wobble" mode 
for background subtractions in the later analysis.

Prior to these observations, we observed this SNR from 2004 January to 
February with two telescopes (T2 and T3).
and preliminary results can be found in \citet{tanimori}.
The statistics of the OFF-source runs were insufficient, only 1/3 of the
ON-source runs. Therefore, we describe that data set in the Appendix 
and concentrate on the description of 2005
observations in the body of this paper.

The light collecting efficiencies, including the reflectivity
of the segmented mirrors, the light guides, and the quantum efficiencies
of photomultiplier tubes were monitored by a muon-ring analysis
\citep{enomoto_vela}. 
The light yield per unit arc-length is approximately proportional
to the light collecting efficiencies.
The ratios of these at the observation period with respect to the
mirror production times (i.e., deterioration factors) 
were estimated to be 45, 55, and 73\% for
T2, T3, and T4, respectively. The measurement errors are considered to
be at less than the 5\% level.
These values were checked analyzing Crab data which were obtained in
2004 November, which are described in the next section.
The deteriorations were mostly due to dirt and dust settling on the
mirrors.
We cleaned the mirrors with water in October 2005 
and the partial improvement (a factor of 1.3-1.4) of
the light collecting efficiencies were observed.


\section{Analysis}

The analysis procedures used were identical with those 
described in \citet{enomoto_vela} for
the Vela pulsar wind nebula, 
which is approximately 0.5 degree south of the Vela Pulsar.
As a full instrumental description was given in
\citet{enomoto_vela}, we omit a detailed discussion here.

At first, the Hillas parameters \citep{hillas} 
were calculated for the three telescopes' images.
The gamma-ray incidence directions were adjusted by minimizing
the sum of squared widths (weighted by the photon yield) 
of the three images seen from the assumed position (fitting parameter).

In order to derive the gamma-ray likeliness,
we used an established mathematical method, 
the Fisher Discriminant (hereafter $FD$ in short) \citep{fisher}.
When we use a multi-parameter set such as:
$$\vec{P}=(W2,W3,W4,L2,L3,L4),$$
where $W2,W3,W4,L2,L3,L4$ are energy corrected $widths$ and $lengths$ for the
T2, T3, and T4
and assume that a linear combination of
$$FD=\vec{\alpha}\cdot\vec{P},$$
provides the 
best separation between signal and background, then the set of
linear coefficients ($\vec{\alpha}$) should be uniquely determined as
$$\vec{\alpha}=\frac{\vec{\mu}_{sig}-\vec{\mu}_{BG}}{E_{sig}+E_{BG}},$$
where $\vec{\mu}$ is a vector of the mean value of $\vec{P}$ for each
sample and $E$ is their correlation matrix.
This method is widely used in the high-energy experiments such
as the $B-factory$ in order to separate spherical events from
jet-like ones \citep{abe}. We previously used 
it in Vela Pulsar analysis \citep{enomoto_vela}
to separate ``sharp'' (gamma-ray--like) images from ``smeared'' (background) 
ones.
The values of
$\vec{\mu}_{sig}$, $\vec{\mu}_{BG}$,
$E_{sig}$, and $E_{BG}$ can be calculated from the
Monte-Carlo and observational data (OFF-source runs), respectively.

\begin{figure}
\plotone{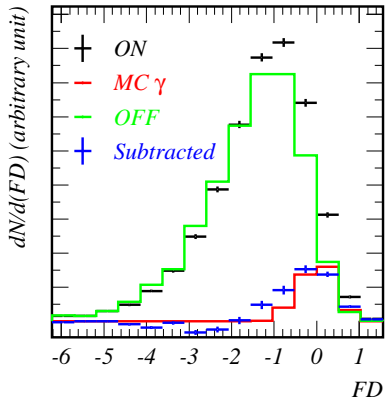}
\caption{
Fisher Discriminant ($FD$) distributions.
The black points with the error bars are those for events
inside a one degree circle centered on the SNR.
The green histogram is for OFF-source data.
The normalization of the latter was determined by the fitting
procedure described in the text.
The blue points show the background subtracted signals
and the red histogram
is the best fit for gamma-rays from Monte-Carlo simulations.
}
\label{ffd2005}
\end{figure}

$FD$ is a linear combination of multi-parameters.
We, therefore, should select parameters which can be linearly added.
The $width$s and $length$s
of T2, T3, and T4 are
second order cumulative moments of shower images
and thus their linear combination is a reasonable
assumption.
In order to keep the mean of the Hillas parameters
(which are energy dependent) at the same values in various energies, we
corrected the energy dependence of each parameter subtracting 
best-fitted second-order polynomials for the Monte-Carlo gamma-rays,
i.e., the means of the corrected Hillas parameters are all set to zero
for gamma-rays, independent of energy.
This procedure ensures the mean position of the $FD$ to be located 
at zero for gamma-rays.
Also a linear combination of six random parameters 
ensures the distribution has an approximately
Gaussian form (see the red histogram in Fig.~\ref{ffd2005}
of Monte-Carlo gamma-ray events).
The contributions of Hillas parameters to the photon--proton separation 
can be expressed by
$\alpha_i\cdot P_i$, which were (0.06, 0.25, 0.48, 0.28, 0.31, 0.16) for
$W2,~W3,~W4,~L2,~L3,~L4$, respectively. 
The main differences of three telescopes are the different
spot sizes of mirrors.
As was expected, the $width$ of T4 gave the
largest contribution which reflects the respective mirror qualities.
The $FD$ can be understood as an approximate representation of
``mean scaled $width$" which was energy corrected and linearly corrected
by the other Hillas parameters.
The advantage of this method is that it can be calculated
fully automatically 
in the analysis procedure, i.e., any IACT can reproduce it.

\begin{figure}
\plotone{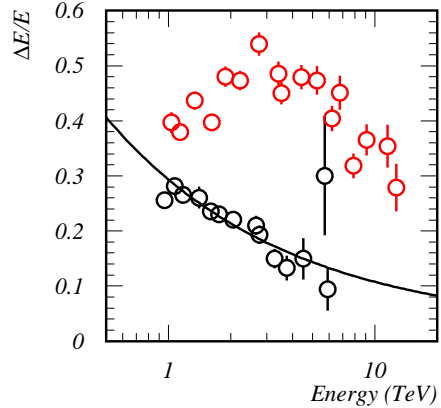}
\caption{
The energy resolution ($\Delta E/E$) estimated by Monte-Carlo
simulations. 
The red points are obtained before the ``edge cut" and the black points
after it.
The curve is the best-fit function described in the
text.
}
\label{fdee}
\end{figure}

Here, we rejected events with hits in the outermost layer of the cameras
(``edge cut"). These rejected events cause finite deformations especially
in the $length$ distribution which results in deformations of the $FD$. 
This cut was 
introduced after the analysis in \citet{enomoto_vela}
and was found to improve the energy resolution significantly:
$\Delta E/E$ was estimated using
Monte-Carlo simulations and is plotted in Fig.~\ref{fdee}.
The red points are those before the ``edge cut" and the black
after. A dramatic improvement in energy resolution is obtained.
It was parametrized by the function
$\Delta E/E=27/\sqrt{E/1~TeV}+2\%$ (the solid curve in Fig.~\ref{fdee}.).
Here, we did not apply either the ``core distance" or ``slant distance"
corrections described in \citet{hofmann}.
The gamma-ray energies were simply estimated by the total
number of photoelectrons observed.
Although this cut significantly
reduces gamma-ray acceptance, especially in the higher energy region,
we decided to use it in order to safely determine the gamma-ray 
energy spectrum in this analysis.
The improvement of acceptance may be considered in more detail
elsewhere in the future.

Since we have $FD$ distributions for OFF-source data and the 
Monte-Carlo gamma-ray
events, we can assume these are background and signal behaviors.
We, therefore, can fit the $FD$ distribution of ON
(the black points in Fig.~\ref{ffd2005}) with the above emulated
signal and real background functions, to derive the number of signal events
passing the selection criteria. 
With this fit, we can determine the gamma-ray excess without any positional
subtractions, i.e., we can analyze all events in the FOV and not only
the events which comes from the direction of the SNR.
This is a two-parameter fitting and these coefficients can be exactly
derived analytically.
To summarize, there are two differences in the analysis procedure
from the Vela paper \citep{enomoto_vela};
\begin{itemize}
  \item events with hits in the outermost layer were rejected (``edge cut"), 
        and
  \item OFF-source events were used for emulating the background
        behavior of the $FD$.
\end{itemize}
In contrast, in the previous analysis, the background function was
made from the ON data (outside of the source region), 
because of the lack of statistics of the OFF data.
The $FD$ has a position dependence and it deforms 
near the edge of camera.
Thus with the high statistics OFF events, 
the current background determination is improved and its
position dependence inside the FOV of the camera
was taken into account.
The result of the fitting is shown by the blue points
(the background-subtracted signal) and the red histogram 
(the best-fit signal)
in Fig.~\ref{ffd2005}, while the green is the best-fit background.
We have a statistically significant signal 
within one degree of the SNR center.
Note that this fitting procedure gives a maximum acceptance
with reasonably small statistical errors without 
introducing any image cuts, 
i.e., a $\chi^2$ fit gives the maximum acceptance of events remaining
after the pre-selections and automatically minimizes the subtraction ranges.
In addition, it removes any subjective bias in cut determination.

\begin{figure}
\plotone{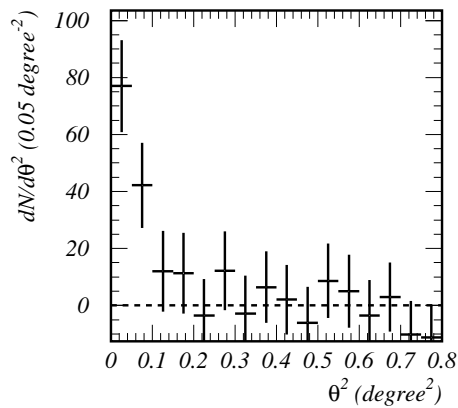}
\caption{
The $\theta^2$ plot for the 2004 November Crab nebula data.
}
\label{ftcrab}
\end{figure}

This method was checked by analysis of Crab nebula data taken in November
2004. The wobble-mode observation was also used. 
The analyzable data corresponded
to 316.4 min. The $\theta^2$ plot is shown in Fig.~\ref{ftcrab}.
The number of excess events is 119 $\pm$ 22 (5.4$\sigma$).
The flux is 1.2$\pm$0.3 times the standard Crab flux with the power-law
consistent with the standard index of $-$2.5.

\begin{figure}
\plotone{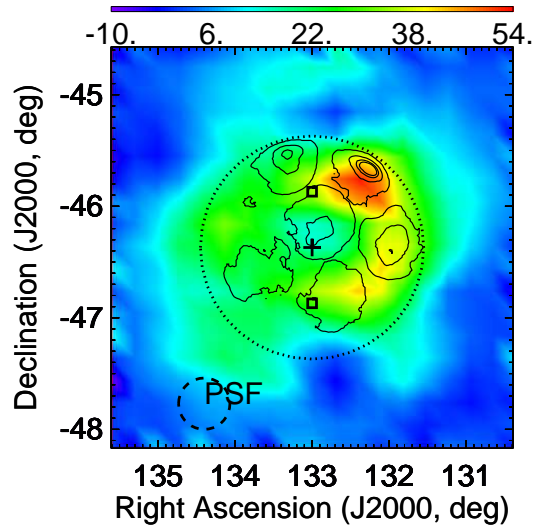}
\vskip 0.5cm
\caption{
Morphology of gamma-ray--like events. 
The smoothing was carried out using the average of the center and
neighboring eight pixel where pixel size was 0.2 $\times$ 0.2 degree$^2$.
The vertical scale (number of
events per pixel) is
indicated in the top bar. 
The dotted circle of 1 degree from the supernova center is shown.
The dashed circle of 0.23 degree (plotted in lower left) 
is the ($1\sigma$) point spread function
, where 68\% of events are contained.
The cross indicates the average pointing position, i.e., the center of
the remnant and the squares the ''wobble" pointing positions.
The contours are the 
20, 45, 65, and 80\% levels in the ASCA Gas Imaging Spectrometer X-ray map
\citep{tsunemi,slane}.
}
\label{fm2005}
\end{figure}

H.E.S.S.\ produced a very clean gamma-ray count map showing the
shell-like structure of RX J0852.0$-$4622.
In making a 2D-count map, we tried a simple time-based subtraction.
The normalization factor for OFF-source run is 1081/1031=1.05.
We made two 2D-plots for arrival directions for ON and OFF
using selected events with the $FD$ in the range between $-$1 and 1,
which is the signal region as shown in Fig.~\ref{ffd2005}.
The difference between two 2D distributions with the above normalization
is shown in Fig.~\ref{fm2005}.
Comparing it with Fig.~2 of \citet{hess0852}, we confirm the shell-like
structure, with less emission in the center position. Our data show
that the emission in the northwest-rim (NW-rim) is the strongest.
The signal-to-noise ratio (S/N) of our analysis is significantly
worse than that of H.E.S.S., which can be seen comparing the S/N of Fig.~1
in \citet{hess0852} and Fig. \ref{ffd2005} of this paper.
Therefore, the statistical error in this map is larger than that in the 
H.E.S.S.\ map.
The difference between red and green of this color map (Fig. \ref{fm2005}) is
not significant. 

\begin{figure}
\plotone{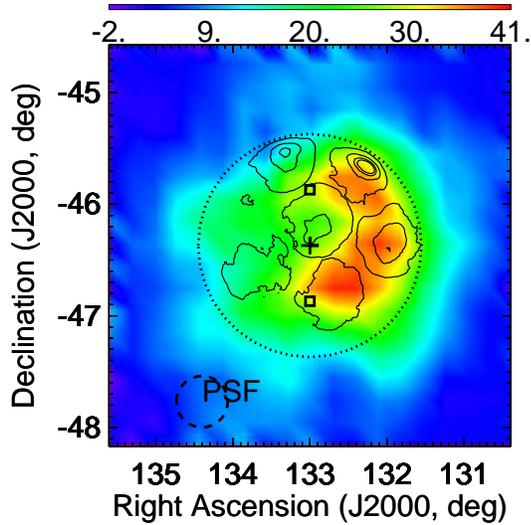}
\vskip 1cm
\caption{
Morphology of gamma-ray--like events obtained by the fitting procedure
described in the text. 
The other definitions are as same as those of Fig.~\ref{fm2005}.
}
\label{fmfit}
\end{figure}

Fig.~\ref{fm2005} was made by positional subtraction.
Fig.~\ref{ffd2005} was made based on fitting the event shapes ($FD$).
We can also study the morphology by extracting gamma-ray--like events
using the fitting procedure described so far.
The $FD$ distributions for each direction are recorded in multiple
histograms and the fittings were carried out.
The resulting morphology is shown in Fig.~\ref{fmfit}.
Although we observe a shell-like structure, 
the brightest region has changed from that of Fig.~\ref{fm2005}.
As has been described, the difference between 
this figure and Fig.~\ref{fm2005} can be considered to be
within the level of statistical fluctuations.

We have thus confirmed the H.E.S.S.\ results that 
RX J0852.0$-$4622
is a strong, shell-like TeV gamma-ray source.
We have also confirmed our previous claim of the CANGAROO-II 
detection of gamma-rays from the bright 
NW-rim (though not the reported soft spectrum).
The threshold of this analysis (Figs.~\ref{ffd2005}, and \ref{fm2005}) 
is estimated to be 1.02\,TeV.
This high threshold can be explained by the deterioration of 
``light collection efficiencies". 
The relatively poor S/N, even for the three-fold coincidence data, can be
explained by the blur spot size of $>$0.1 degree of the segmented
mirrors, which are made of plastic.
Monte-Carlo simulations with zero spot size show a much better
separation between gamma-ray and proton events.
This means that replacing these mirrors with glass or metal, we have a
chance to 
approach
H.E.S.S.\ quality even with 10\,m diameter structure of
the present telescopes.

\section{Result}

\begin{figure}
\plotone{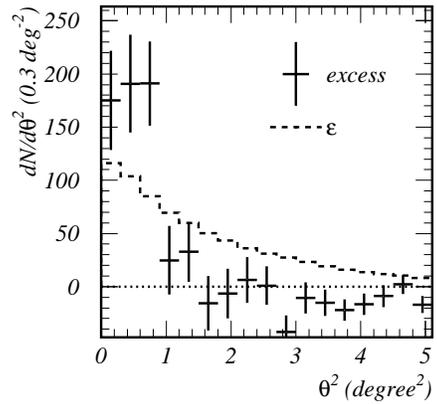}
\caption{
Wide-range $\theta^2$ plot.
The vertical scales are the numbers of excess events per 0.3 degree$^2$.
The dashed histogram is the acceptance (in arbitrary units).
}
\label{ft2005}
\end{figure}

The radial distribution of the gamma-ray signal
with respect to the SNR center is shown in Fig.~1 of \citet{hess0852}.
We took the $FD$ distributions of 
annuli of various squared-radii with respect
to the SNR center and fitted them. 
The $\theta^2$ distribution over a wide range is shown in Fig.~\ref{ft2005}.
The spatial excess 
appears only within 
one degree from the center of the SNR,
in agreement with the H.E.S.S.\ result.
The number of excess events is 557 $\pm$ 77 (7.2$\sigma$).

\begin{figure}
\plotone{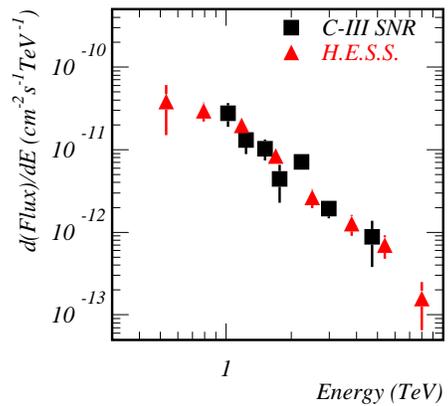}
\caption{
Differential energy spectra;
the red points by H.E.S.S.\ are for the whole remnant
and the black points from these CANGAROO-III observations
are also for the whole remnant.
The error bars are statistical.
}
\label{ffl2005}
\end{figure}

The gamma-ray fluxes within a one-degree radius 
were derived by fitting the $FD$ distributions within
this area on an energy bin-by-bin basis.
The black squares in Fig.~\ref{ffl2005} are obtained by this experiment.
The numerical data are listed in Table~\ref{table1}.
\begin{table}
\begin{center}
\caption{Differential flux within one degree from
the SNR center.}
\label{table1}
\begin{tabular}{ccr}
\hline \hline
Energy binning (TeV) & Mean Energy (TeV) 
& dF/dE (cm$^{-2}$s$^{-1}$TeV$^{-1}$) \\
\hline
0.83 -- 1.07 & 1.02 &( 2.77 $\pm$  0.88) $\times 10^{-11}$ \\
1.07 -- 1.37 & 1.24 &( 1.32 $\pm$  0.43) $\times 10^{-11}$ \\
1.37 -- 1.62 & 1.51 &( 1.03 $\pm$  0.29) $\times 10^{-11}$ \\
1.62 -- 1.91 & 1.77 &( 4.40 $\pm$  2.14) $\times 10^{-12}$ \\
1.91 -- 2.66 & 2.24 &( 7.11 $\pm$  1.17) $\times 10^{-12}$ \\
2.66 -- 3.70 & 2.98 &( 1.93 $\pm$  0.46) $\times 10^{-12}$ \\
3.70 -- 6.07 & 4.72 &( 8.83 $\pm$  5.03) $\times 10^{-13}$ \\
\hline \hline
\end{tabular}
\end{center}
\end{table}
The consideration on the experimental energy resolution was carried out
unfolding the generated energy from the detected photon-electron yield
using the Monte-Carlo events under the assumption that the energy spectrum
is proportional to $E^{-2.1}$ (H.E.S.S.'s spectrum).
The ''Mean Energy" quoted in Table \ref{table1} is the mean generated
energy of the accepted events in the Monte-Carlo simulation in each
binning.
The obtained differential flux is calculated at this ''Mean Energy"
under the assumption of the above power-law spectrum.
The best fit with a power-law assumption for this energy spectrum is:
\begin{eqnarray*}
\frac{dF}{dE}&=&[2.5\pm 0.6(stat.)\pm 0.6 (sys.)]\times 10^{-11}\\
             & &\cdot
\left(\frac{E}{1~{\rm TeV}}\right)^{2.2\pm 0.3(stat.)\pm 0.3(sys.)}
~[{\rm cm}^{-2}{\rm s}^{-1}{\rm TeV}^{-1}].
\end{eqnarray*}
Presently we estimated the systematic errors as follows;
the main one is the energy scale of 15\% due to the absolute light collection
efficiency and the rest is the detector systematics of 10\%.
The ambiguities of the absolute flux and the power-law strongly correlated
to each other.
The data points of \citet{hess0852} were obtained from 
$$\rm{http://www.mpi-hd.mpg.de/hfm/HESS/HESS.html}$$
Our results are consistent with H.E.S.S.\ for both flux level and spectrum,
i.e., the CANGAROO-III results for the whole remnant yield
a harder spectrum than the CANGAROO-II spectrum of \citet{katagiri}.

\begin{figure}
\plotone{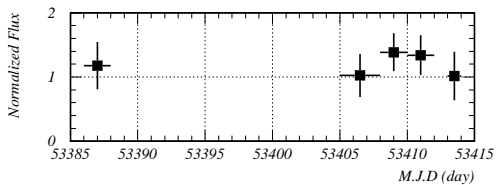}
\caption{
Time variability of the integral fluxes at the energy greater than 1.02\,TeV. 
The vertical axis is the flux normalized
to the H.E.S.S.\ mean flux 
($dF/dE=2.1 \times 10^{-11}E^{-2.1}$ [cm$^{-2}$s$^{-1}$TeV$^{-1}$]).
The horizontal axis is modified Julian days (M.J.D.).
}
\label{frun}
\end{figure}

We checked the time variability of these excesses. Five periods
were selected so that they have similar statistics,
i.e., two-to-four days periods.
The integral fluxes at energy greater than 1.02\,TeV, 
which are normalized to the H.E.S.S.\ mean flux
are plotted in Fig.~\ref{frun}.
It is consistent with stable emission.

\begin{figure}
\plotone{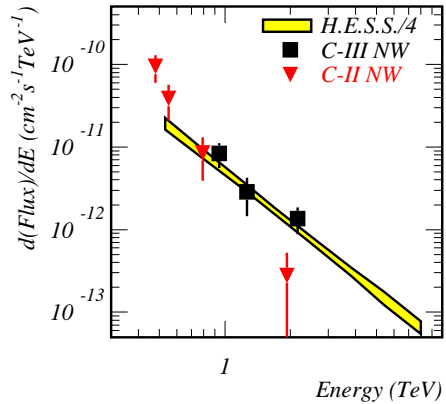}
\caption{
Differential energy spectra;
the red points are obtained by CANGAROO-II around the NW-rim
and the black points
are obtained by this observation for the quadrant centered around
the NW-rim.
The yellow area is 1/4 of the H.E.S.S.\ flux.
}
\label{fflnw}
\end{figure}

We now consider the fluxes and spectra as a function of position
in the SNR to attempt to reconcile the differences between 
the CANGAROO-II and H.E.S.S.\ results.
The energy spectrum obtained by CANGAROO-II  
is shown by the red points
in Fig.~\ref{fflnw}.
The CANGAROO-II flux is barely consistent with the other results.
The only difference is that
the single telescope CANGAROO-II observations
yielded an excess only at the NW-rim.
The exact procedure is reviewed as follows;
\begin{itemize}
  \item Derive Hillas parameters and calculate $Likelihood~Ratio~(LR)$.
Probability Density Functions (PDFs) were derived for
both gamma-ray and cosmic ray initiated events.
The PDFs for gamma-rays were obtained from simulations,
while those for cosmic rays were obtained from OFF-source data,
both in energy-by-energy bases. 
Histograms were made of  \textit{length} and \textit{width} 
using both data sets; these distributions were then normalized to unity. 
The probability (L) for each assumption was thus obtained by multiplying 
PDF(\textit{width}) by PDF(\textit{length}). 
In order to obtain a single parameter, and also to normalize it to unity, 
we used the Likelihood-ratio ($LR$):
$\frac{L(gamma-ray)}{L(gamma-ray)+L(proton)},$
therefore, in energy-by-energy bases.
  \item Select $LR>0.4$ events and plot $Alpha$(``Image oriented angle"). 
  \item Normalize ON- and OFF-source distributions 
     at $Alpha>30$ degrees and obtain the
     excess within $Alpha<30$ degree. We did not use the time-based
     normalization (considering the ON- and OFF-source observation
     time ratios). Only the shape of the
     OFF-source $Alpha$ distribution was used 
     to determine the normalization factor.
\end{itemize}
Only in the case that the spectral index in the NW-rim is softer
than elsewhere 
can the softer spectrum obtained by CANGAROO-II be explained.
We counted the excess within the quadrant within $\pm$45 degree from the
NW-rim center (RA, dec = 132.245$^\circ$, $-$45.650$^\circ$ J2000).
The result is shown by the black points in Fig.~\ref{fflnw}
which is slightly higher than 1/4 of the whole emission
(the hatched area in Fig.~\ref{fflnw}). Its spectrum is still consistent with
that of the whole remnant. 
Therefore,
it is not consistent with the CANGAROO-II flux within the statistical errors.
Additional systematic uncertainties such as outlined in \citet{katagiri}
are required.
The main one was considered to be energy uncertainty due to the deterioration
of mirrors and PMTs. Also saturation of the electronics might contribute.
In particular, the lower energy fluxes were derived with large acceptance
correction factors and have larger systematic errors.

\begin{figure}
\plotone{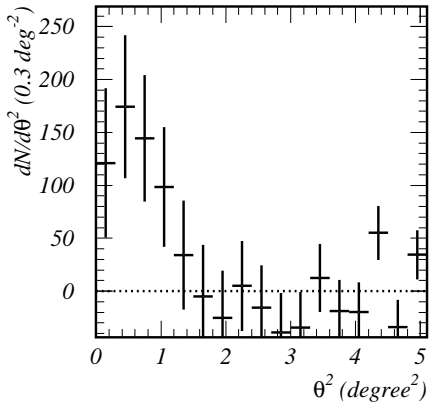}
\caption{
Wide range $\theta^2$ plot obtained using the ``Likelihood analysis".
}
\label{ftwatanabe}
\end{figure}

\begin{figure}
\plotone{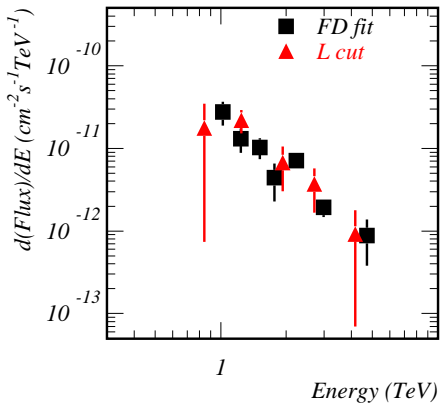}
\caption{
Differential energy spectra;
the red points were obtained using the ``Likelihood-Ratio analysis" 
for the whole remnant 
and the black points by the $FD$ analysis
for the the whole remnant. 
The error bars are statistical.
}
\label{fflwatanabe}
\end{figure}

We analyzed the same data again using a likelihood-ratio 
instead of the Fisher Discriminant for the gamma/hadron separation
in order to check the CANGAROO-II analysis.
Here we used $lengths$ and $widths$ of three telescopes and used
a safety cut of $LR>0.1$
for this supernova remnant
in Fig.~\ref{ftwatanabe}.
We see a similar excess within the remnant. 
The number of excess events is 
538 $\pm$ 128 (4.2$\sigma$),
consistent with that obtained
by the $FD$ analysis.
The flux obtained by this method is shown in Fig.~\ref{fflwatanabe}.
The best fit with a power-law assumption is:
$\frac{dF}{dE}=[2.4\pm 0.8(stat.)\pm 0.6 (sys.)]\times 10^{-11}\cdot
\left(\frac{E}{1~{\rm TeV}}\right)^{2.0\pm 0.4(stat.)\pm 0.3(sys.)}
~{\rm cm}^{-2}{\rm s}^{-1}{\rm TeV}^{-1}.$
Therefore, the detection of this supernova remnant
by \citet{katagiri} is shown to be correct.
Hardware correction for the ``light collecting" efficiencies and 
``saturation" effect have been refined since the CANGAROO-II results,
and since the CANGAROO-II telescope (T1) has not been operated 
in recent years,
it is hard to estimate these deteriorations over 2001 to 2003.

\section{Discussion}

We now discuss the possible origins of these TeV gamma rays.
The key constraints are the energy of the protonic cosmic rays 
are $\sim 10^{50}$ and the electronic $\sim 10^{48}$ erg.
The former is required for the supernova origin of cosmic rays
and the ratio between the former and latter is based on the electron
measurements in the solar system \citep{kobayashi}.
We, here, argue the distance of this SNR and 
the mechanism of the gamma-ray emission.

We first modeled the TeV gamma-ray flux assuming
Inverse Compton scattering on the cosmic microwave background. 
The input electron spectrum is assumed to be 
proportional to $E^{-\gamma} e^{-E/E_{max}}$.
Here the number of the data points are 16 (H.E.S.S.=9 and CANGAROO-III=7).
The best-fit results under a fixed parameter of $\gamma=2.1$ 
(a typical assumption) are 
for an exponential cut-off energy
of $E_{max}=26\pm 5$\,TeV with $\chi^2/dof$=13.9/(16$-$2).
Under the assumption of d=200\,pc (distance of this SNR from earth)
we obtained total electron energy at $E$ greater than 1\,GeV 
of $E_e=0.04\pm 0.01\cdot 10^{48}$\,erg
which is low considering the supernova origin of 
cosmic rays under the assumption of the electron/proton ratio (e/p) 
at the acceleration site \citep{reynolds,kobayashi}. 
Note that we only consider the statistical errors of the observation. The
real errors of the fitted parameters would be larger than those cited.

Second we consider protonic contributions to the TeV flux.
Assuming a shock velocity of 5000\,km\,s$^{-1}$ in the free expansion era
and a swept mass of order $M_\odot$ with the age of this SNR of 
630\,yr (d=200\,pc), the ambient interstellar matter (ISM) 
density of n=0.2\,p/cc
is obtained. In the case of d=1\,kpc, n=0.002\,p/cc.
We carried out a two-parameter fit ($E_{max}$ and $E_p$) to only the TeV flux 
with a fixed parameter of $\gamma=2.1$
where the input proton flux was assumed again to be proportional to
$E^{-\gamma} e^{-E/E_{max}}$.
The best fit was obtained for $E_{max}>70$\,TeV 
with $\chi^2/dof=15.2/(16-2)$.
The protonic CR energy at d=200\,pc is 
$E_p=0.80\pm 0.08 \cdot 10^{50}$\,erg under
the assumption of the ambient ISM density of
0.2\,p/cc at d=200pc.
We can not reject a protonic contribution to the TeV emission
for d=200\,pc. 
We, however, should have a significant electronic contribution and
to produce a significant protonic component we need to suppress the
e/p-ratio to far less than 0.01.
When we constrain the total cosmic-ray energy to be less than $10^{50}$\,erg
at d=1kpc,
the ambient density of the ISM is required to be at least 4\,p/cc.
A total mass of
$\sim550\frac{\rm n}{1 \rm{p/cc}}M_\odot$ within a 1 degree radius
is necessary in that volume with a shell-like distribution. 
This kind of heavy of molecular cloud is, however, 
marginally within the detectable
range of recent CO observations.
The search for such a molecular cloud is necessary.

We now shift to the lower energy region.
The radio data used here are based on Table~2 of \citet{duncan} and
the X-ray data are from Table~1 of \citet{slane} (ASCA).
Note that \citet{slane} show fluxes for the north-west, north-east, and west
regions of this SNR. We combined them and calculate five data points from
0.5 to 10\,keV. They only gave errors for the spectral indices. We, therefore,
take only them into account, i.e., the errors are underestimated.
With the above best-fit parameters under the electron assumption, 
an ambient magnetic field of 
$B=\sim 4~\mu$G also gives an allowable one-parameter fit to the
X-ray spectrum and lower estimation to the radio observation.

\begin{figure}
  \plotone{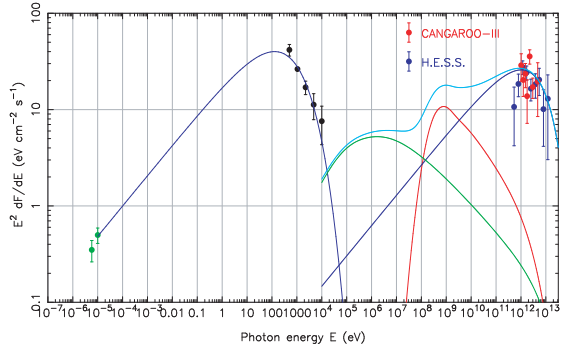}
  \caption{
  Spectral energy distributions. The data points are for radio, X-ray, 
  H.E.S.S., and this observation. The colored curves are model 
  calculations; the blue at 
  low energy are for synchrotron radiation and at high energy 
  are Inverse Compton,
  the red are $\pi^0$ decay, the green, Bremsstrahlung, and the light blue 
  show the sum at high energy.
  }
  \label{fsed}
\end{figure}

A filamental X-ray structure was observed for this SNR by Chandra
\citep{bamba}.
We might need to introduce a magnetic field filling factor ($f_B<1$).
The best-fit results freeing all five parameters are shown in Fig.~\ref{fsed}.
The $\chi^2/dof$ was 18.8/(23$-$5) with 23 data points.
The fitted spectral energy distribution is shown in Fig.~\ref{fsed}.
The blue lines are the electronic contributions described above.
We obtained: $B=5.8\pm 0.3\,\mu$G, $f_B=0.40\pm 0.04$,
$\gamma=2.37\pm 0.01$, $E_{max}=37\pm 2$\,TeV, and $E_e=0.18\pm 0.02~\times
10^{48}$\,erg at d=200\,pc, less energetic than that of 
the standard supernova origin.
Also the filling factor looks too large to create the X-ray 
filamental structure.
If this SNR is assumed to be located at d=1\,kpc, 
the energy simply becomes
25 times bigger, that is, marginally higher than usually assumed.
This low magnetic field strength with the estimated age of 630\,yr at
d=200\,pc is 
considered to be within the
allowable range for theory of electron acceleration up to 40\,TeV. 
Assuming the proton spectrum is the same as that of the electrons
and its energy is hundred times larger,
the red line is the
expectation of the $\pi^0 \rightarrow \gamma\gamma$ contribution
which is a negligibly small contribution to the TeV emission.
The green line is that for Bremsstrahlung. Note that this calculation
has a big uncertainty in the sub-MeV region.
Suzaku observations of the hard-X region are eagerly awaited.
In the case of $\gamma=2.1$, 
the contribution from Bremsstrahlung is below the Suzaku
sensitivity limit.

Considering the filamental structure observed in X-rays,
\citet{bamba} assumed a magnetic field as high as 500\,$\mu$G.
This model predicted the maximum energy of the electrons 
to be $E_{max}=3$\,TeV.
With that energy spectrum, the observed flat
TeV emission can not be well described via either
Inverse Compton or $\pi^0$ decay processes.
In order to fit the TeV spectrum, 
an order of magnitude higher cut-off energy is definitely necessary,
which favors a two orders of magnitude
lower magnetic field, because the maximum energy
of the synchrotron radiation is a function of $E_{max}\sqrt{B}$.
This contradiction can be resolved by introducing a different
energy spectrum between electrons and protons as was predicted
by kinetic non-linear acceleration theories where synchrotron
cooling of electrons was taken into account (see for example \citet{bkv}).
The filamental structure might be a result of strong
synchrotron losses of emitting electrons in amplified fields as in case
of other SNRs \citep{vink,bkv,bv}.
Such a high magnetic field (hundreds of $\mu$G) in the downstream region
reject an Inverse Compton solution thus supporting the idea of
a nucleonic origin of the observed TeV gamma-ray flux.
However, in that case, 
we need an origin other than SNRs for the observed electrons 
in the solar system.
Our simple estimation agrees with their models \citep{bkv}.

Although at present a perfect interpretation is missing,
fine angular resolution studies
over a broad energy range should help to determine one. 

\section{Conclusion}

We have observed the supernova remnant RX J0852.0$-$4622 with the
CANGAROO-III stereoscopic imaging Cherenkov telescope.
It is clear from the results for RX J0852.0$-$4622 that
CANGAROO-III's stereoscopic performance is significantly 
improved from that of the single CANGAROO-II telescope. 
We have confirmed the H.E.S.S.\ group's result 
about the morphology and the energy spectrum.
The NW rim is the brightest feature.
RX J0852.0$-$4622  has been confirmed to be one of the strongest 
TeV gamma-ray sources
in the southern hemisphere and shows a clear shell-type 
structure at these energies.
The energy spectrum around the NW-rim was measured to be
consistent with that of the whole remnant, not in good
agreement with the previous CANGAROO-II result.
The difference can be partially explained by the deterioration
of the hardware of the CANGAROO-II telescope.

\section*{Acknowledgments}

This work was supported by a Grant-in-Aid for Scientific Research by
the Japan Ministry of Education, Culture, Sports, Science and Technology, 
the Australian Research Council, JSPS Research Fellowships,
and Inter-University Researches Program 
by the Institute for Cosmic Ray Research.
We thank the Defense Support Center Woomera and BAE Systems.

\section*{Appendix: 2004 observation results}

Additional observations were carried out 
in the period from 2004 January 17 to February 25
in the same ``wobble mode" as the 2005 observations.
The average pointing position was the NW-rim (RA, dec = 132.245$^\circ$,
$-$45.650$^\circ$ J2000), i.e., the  maximum of the X-ray emission.
However, only the second and third telescopes, T2 and T3, were used for the
2004 observations as
T4 became operational
after these observations had been completed.
Also in this period the ``Global Trigger" system was not ready.
Data were recorded for T2 or T3 independently when
the photomultiplier signals exceeded 7.6 photoelectrons (p.e.) 
in more than four 
camera pixels (``Local Trigger").
The GPS time stamp was recorded in each dataset. 
An offline coincidence of time stamps from T2 and T3 within
$\pm$100\,$\mu$s \citep{enomoto_vela} was required for a stereo event.
The typical trigger rate for each telescope was 80\,Hz, which was
reduced to 12\,Hz for stereo events.
The light-collecting efficiency was calibrated during the offline processing
by analyzing $\mu$-rings 
and found to be 70$\pm$5\% for both T2 and T3,
as for the Vela Pulsar
observations \citep{enomoto_vela}.
The total observation time was 1781, and 436~min, for ON- and OFF-source
data, respectively.
After pre-selection, the event rate was reduced to 10\,Hz.
Finally after the ``cloud and elevation cut''
the analyzable data (ON-source run) corresponded to 1120 min.\, with a
mean elevation angle of 70.3$^\circ$.
The OFF-source data, however, only had 379 min., in part as
these observations were carried out at a similar time with the
CANGAROO-III observations of the Vela pulsar region described by
\citet{enomoto_vela}.

The differences between these and 2005 data are summarized as follows;
\begin{itemize}
\item 2-fold (software) coincidence in 2004 and 3-fold (hardware) coincidence in 2005, and
\item the OFF-source data in 2004 are only 1/3 of 
the ON-source data,
\end{itemize}
Although the light collecting efficiency is higher than 2005,
there remain some uncertainties, especially in the background subtraction.
We, therefore, report these results in this Appendix as supporting evidence 
to the main arguments.

The same kind of analysis as for the 2005 data were carried out.
Here, we restrict the analysis only to the higher acceptance regions
due to the lack of statistics in OFF-source runs.
The higher acceptance region was chosen as within 1.4 degrees from
each ``wobble" pointing. Only the overlapping region
was analyzed. This is approximately the north-east half of the remnant.
Comparing with the 3-fold data, the signal-to-noise ratio
is poor. The 3-fold coincidence data had better background
rejection power
than that of the 2-fold data.
The errors in the subtracted data are dominated by the OFF-source
statistical errors.
Due to these, the statistically accurate 
background functions could not be obtained
position-by-position bases in the FOV.
The background also behaved non-uniformly near the edge of the camera.
Therefore, we did not use the fitting procedure of the
2005 analysis. The ON$-$OFF subtraction was carried out using time-based
normalization factor.

\begin{figure}
\plotone{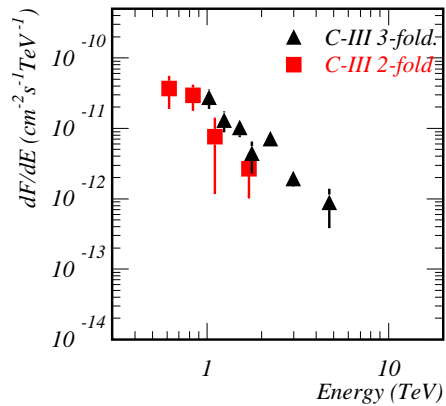}
\caption{
Differential energy spectra;
the black points from the 3-fold coincidence analysis of
2005 CANGAROO-III data for the whole remnant (the same as the black points
in Fig.~\ref{ffl2005}, and 
the red points from the 2-fold analysis of 2004 CANGAROO-III 
for about a half of the remnant.
}
\label{ffl2004}
\end{figure}

As has been described, we can only analyze half of the remnant.
The flux inside the maximum acceptance region is shown in 
Fig.~\ref{ffl2004},
i.e., fluxes of the half remnant.
The energy resolution was estimated using Monte-Carlo simulations
in the same way 
as the 2005 analysis to be $\Delta E/E=17/\sqrt{E/1~TeV}+12\%$, which is
slightly worse than the 3-fold analysis.
The energy spectrum is similar and the flux is half compared to the
whole remnant.
This is supporting evidence for the results of the 2005 observations.

\end{document}